\documentclass[prb,showpacs,showpacs,floatfix,twocolumn]{revtex4}
\usepackage{graphicx}

\begin{document}

\newcommand*{\cm}{cm$^{-1}$\,}

\title{Optical properties of pyrochlore oxide $Pb_{2}Ru_{2}O_{7-{\delta}}$}

\author{P. Zheng}

\author{N. L. Wang}

\author{J. L. Luo}

\affiliation{Institute of Physics and Center for Condensed Matter
Physics, Chinese Academy of Sciences, P.~O.~Box 603, Beijing
100080, P.~R.~China}%

\author{R. Jin}
\author{D. Mandrus}

\affiliation{Condensed Matter Sciences Division, Oak Ridge
National Laboratory,
Oak Ridge, TN 37831}%


\begin{abstract}
We present optical conductivity spectra for
$Pb_{2}Ru_{2}O_{7-{\delta}}$ single crystal at different
temperatures. Among reported pyrochlore ruthenates, this compound
exhibits metallic behavior in a wide temperature range and has the
least resistivity. At low frequencies, the optical spectra show
typical Drude responses, but with a knee feature around 1000 \cm.
Above 20000 \cm, a broad absorption feature is observed. Our
analysis suggests that the low frequency responses can be
understood from two Drude components arising from the partially
filled Ru $t_{2g}$ bands with different plasma frequencies and
scattering rates. The high frequency broad absorption may be
contributed by two interband transitions: from occupied Ru
$t_{2g}$ states to empty $e_{g}$ bands and from the fully filled O
2p bands to unoccupied Ru $t_{2g}$ states.
\end{abstract}


\maketitle

Pyrochlore compounds with general formula $A_{2}B_{2}O_{7}$ are
face-centered-cubic oxides with space group $Fd\overline{3}m$ (A
and B are cations). While B cation is six-fold coordinated and
locates at the center of the distorted octahedra formed by corner
O ions denoted as O(1), A cation is eight-fold coordinated with
six O(1) and two other oxygen ions O(2). The $BO_{6}$ octahedra
are corner-sharing and compose a three-dimensional tetrahedral
network, namely, the pyrochlore lattice. Intensive investigations
of pyrochlore oxides have revealed a remarkable range of
interesting and complex phenomena including colossal
magnetoresistive effect, heavy Fermion behavior,
superconductivity, spin ice, spin glass and metal-insulator
transition by selecting different A and B cations. Among those
phenomena, the metal-insulator(MI) transition of ruthenates (B =
Ru) is unexpected and has thus attracted much attention
\cite{Ishii,Subra,Takeda,Kobayashi,Yoshii,Kennedy,Lee,Cox}.

In general, the electron correlation degree of 4d materials is
smaller than that of 3d materials and is thought to be within the
intermediate-coupling regime. For pyrochlore ruthenates,
$A_{2}Ru_{2}O_{7}$, the electrical properties show systematic
change from a Mott insulator to a metal depending on A cation. For
example, $Y_{2}Ru_{2}O_{7}$ is an insulator \cite{Subra};
$Tl_{2}Ru_{2}O_{7}$ exhibits a metal-insulator transition at 120
K,\cite{Takeda} accompanied with a structural change from cubic to
orthorhombic symmetry. Both $Pb_{2}Ru_{2}O_{7-{\delta}}$
\cite{Kobayashi} and $Bi_{2}Ru_{2}O_{7}$ \cite{Yoshii} remain
metallic electrical properties from room temperature to lowest
measured temperature. It is found that metallic $A_{2}Ru_{2}O_{7}$
has a greater Ru-O-Ru bond angle than those of insulating
compounds\cite{Kennedy2}. The angle is affected by A cation. There
are two possible roles of A cation: (1) modifing the Ru 4d band
width through the change of Ru-O-Ru bond angle\cite{Cox}, and (2)
contributing some states to the total states near the Fermi level
\cite{Ishii,Kennedy}.

Optical spectroscopy is a powerful tool to probe the electronic
structure and charge dynamics of a material. Several optical
measurements have been performed on $A_{2}Ru_{2}O_{7}$ compounds,
which yield information about the effect of $\it {A}$ cations on
electronic structure. For $Tl_{2}Ru_{2}O_{7}$, a peak-like feature
in the mid-infrared (IR) region was observed in the optical
conductivity spectra\cite{Lee}, which shifts to low frequencies
with decreasing temperature. For $Bi_{2}Ru_{2}O_{7}$, the mid-IR
feature moves to low frequency region further and overlaps with a
sharp Drude component. In contrast, no mid-IR feature was observed
in optical spectra of insulator $Y_{2}Ru_{2}O_{7}$. Thus, it
appears that the mid-IR feature is related to metallic nature of
$A_{2}Ru_{2}O_{7}$. However, the origin of the mid-IR peak remains
unclear. For $Tl_{2}Ru_{2}O_{7}$, the mid-IR peak is attributed to
the interband transition between the lower Hubbard band(LHB) of Ru
4d $t_{2g}$ energy level and the newly formed midgap state near
the Fermi level caused by self-doping from $Tl_{2}O$ to
$RuO_{6}$\cite{Lee}. For $Bi_{2}Ru_{2}O_{7}$, the
electron-electron correlation is quite weak and its $t_{2g}$
energy level could not split into LHB and UHB bands\cite{Lee2}.
One may thus search for an alternative interpretation for metallic
$A_{2}Ru_{2}O_{7}$.

$Pb_{2}Ru_{2}O_{7-{\delta}}$ is also a Pauli paramagnetic metal in
the whole measured temperature range\cite{Kobayashi}. Among
reported pyrochlore ruthenates, this compound has the least
resistivity. Its resistivity decreases significantly with
decreasing temperature, whereas the resistivity of
$Bi_{2}Ru_{2}O_{7}$ keeps about 600 $\mu\Omega cm$ in the
temperature range between 10 to 300 K.\cite{Yoshii}
High-resolution electron-energy-loss spectroscopy(HREELS) analysis
\cite{Cox} shows that the density of states (DOS) at $E_{F}$ of
$Pb_{2}Ru_{2}O_{7-{\delta}}$ is higher than that of
$Bi_{2}Ru_{2}O_{7}$. All the above experimental phenomena imply
that $Pb_{2}Ru_{2}O_{7-{\delta}}$ is a better metal than
$Bi_{2}Ru_{2}O_{7}$. In this work, we report the optical
conductivity spectra of $Pb_{2}Ru_{2}O_{7-{\delta}}$ single
crystal at different temperatures. At low frequencies, the optical
spectra show typical Drude responses, but with a knee feature
around 1000 \cm. Above 20000 \cm, a broad absorption feature is
observed. Our analysis suggests that the low-frequency responses
may be understood from two Drude components arising from the
partially filled Ru $t_{2g}$ bands with different plasma
frequencies and scattering rates. The high-frequency broad
absorption may be due to interband transitions from occupied Ru
$t_{2g}$ states to empty $e_{g}$ bands and from the occupied O 2p
states to unoccupied Ru $t_{2g}$ bands.

Single crystals of $Pb_{2}Ru_{2}O_{7-{\delta}}$ were grown using a
vapor transport method described in detail elsewhere.\cite{Jin}
The temperature dependent resistivity measured by standard four
probe method is shown in Fig. 1, which indicates a typical
metallic behavior. The resistivity values are lower than the
reported data,\cite{Kobayashi} reflecting high quality of the
crystal. We have measured the frequency dependent reflectivity
R($\omega$) from 50 \cm to 30000 \cm at different temperatures.
The measurements were performed on a Bruker 66v/s spectrometer
with a He flowing cryostat. An \textit{in-situ} overcoating
technique is used for the reflectance measurement.\cite{Homes}
Standard Kramers-Kronig transformations are employed to derive the
frequency-dependent optical conductivity.

\begin{figure}[t]
\includegraphics[keepaspectratio=true, totalheight =2.2 in, width = 3 in]{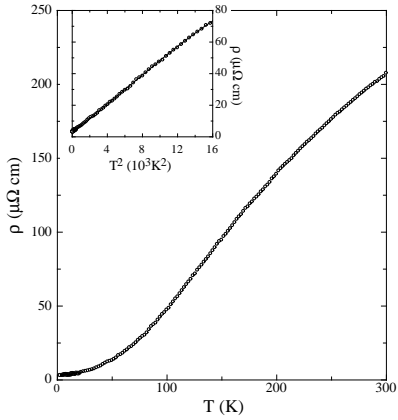}
\caption{Temperature dependence of the electrical resistivity
$\rho$ between 2 and 300 K. The inset is the plot of $\rho$ vs.
T$^2$ between 2 and 120 K. Note the linear behavior of $\rho$ with
T$^2$. }
\end{figure}

\begin{figure}[t]
\centerline{\includegraphics[width=3.0in]{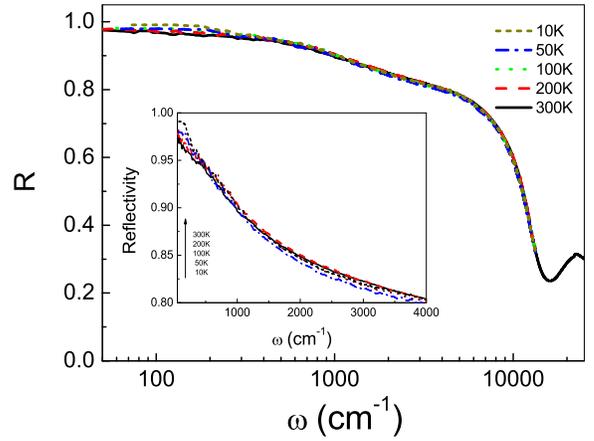}}%
\vspace*{0cm}%
\caption{The plot of frequency-dependent reflectivities at 300 K,
200 K, 100 K, 50 K and 10 K. Inset shows
the low frequency data in an expanded scale.}%
\label{fig2}
\end{figure}

Plotted in Fig. 2 is the reflectivity spectra at various
temperatures. In low-frequency region, the reflectivity
$R(\omega$) slightly increases with decreasing temperature. At
higher but below $8000 cm^{-1}$ frequency region, the reflectivity
slightly decreases with decreasing temperature. $R(\omega$) at
different temperatures cross between 1000 \cm and 2000 \cm. A
plasma edge minimum can be seen at frequency close to $13000
cm^{-1}$.

Fig. 3 is a collection of the real part of the optical
conductivity between 10 and 300 K. In the high frequency side,
there is a broad interband transition peak at about $23000
cm^{-1}$. Peaks at similar energies have been observed in
$Bi_{2}Ru_{2}O_{7}$ and $Tl_{2}Ru_{2}O_{7}$, and were attributed
to the interband transition from filled O 2p band to unoccupied Ru
4d $t_{2g}$ band. Since the transition from occupied $t_{2g}$ band
to empty Ru $e_{g}$ band is also close to this energy\cite{Hsu},
this broad feature may compose of both interband transitions.
Below $13000cm^{-1}$, $\sigma_{1}(\omega)$ shows a Drude-like
response, but with a knee feature near 1000 \cm. Its spectra
weight at about 1000 \cm decreases slightly when temperature
decreases. In $Bi_{2}Ru_{2}O_{7}$\cite{Lee3}, similar absorbtion
feature at 292K is observed too, which is considered as a mid-IR
peak of interband transition inherited from $Tl_{2}Ru_{2}O_{7}$.

\begin{figure}
\centerline{\includegraphics[width=3.0in]{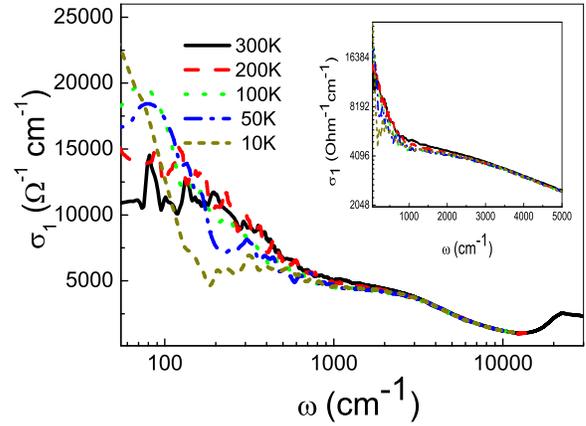}}%
\vspace*{0cm}%
\caption{The plot of frequency dependent conductivities at 300 K,
200 K, 100 K, 50 K and 10 K. Inset shows the low frequency data in
an expanded scale.}
\end{figure}

The sum of the optical conductivity spectral weight gives a
measure of the effective carrier number

$\frac{m}{m^{*}}N_{eff}(\omega_{c})$ of
$Pb_{2}Ru_{2}O_{7-{\delta}}$:
\begin{equation}
  \frac{m}{m^{*}}N_{eff}(\omega_{c})=\frac{2m_{0}}{\pi{e^{2}}N}\int_{0}^{\omega_{c}}\sigma(\omega)d\omega
\end{equation}
in which m is the bare electron mass, $m^{*}/m$ is the ratio of
the effective mass to free electron mass, N is the number of Ru
ions per unit volume and $\omega_{c}$ is a cutoff frequency.
Assuming that $\omega_{c}=1.5 eV$(about $12000 cm^{-1}$) as for
$Tl_{2}Ru_{2}O_{7}$ \cite{Lee}, we get
$\frac{m}{m^{*}}N_{eff}(1.5eV)=0.934$. This value is considerably
higher than that of $Tl_{2}Ru_{2}O_{7}$. That is consistent with
the fact that $Pb_{2}Ru_{2}O_{7-{\delta}}$ is more metallic than
$Tl_{2}Ru_{2}O_{7}$.

In optical conductivity spectra of $Tl_{2}Ru_{2}O_{7}$, there is a
mid-IR peak centering at $4000 cm^{-1}$. \cite{Lee} Band structure
calculations \cite{Ishii} show that, antibonding states of Tl 6s
and O(2) 2p lie in the energy range  from -1eV to 2eV and partly
hybridize with the Ru 4d ($t_{2g}$)-O(1) 2p antibonding states.
That will lead to net charge transfer from $Tl_{2}O(2)$ chain to
the net-work of $RuO(1)_{6}$ octahedra. Such a charge transfer
effect will generate some midgap states within the gap between up
Hubbard band(UHB) and low Hubbard band(LHB). \cite{Lee,Meinders}.
When the midgap states lie above $E_{F}$ and close to LHB, an
interband transition between LHB and those midgap states is
possible. Then a Lorentz-like peak appears in the optical
conductivity spectra. The previous explanation of the mid-IR
absorption feature seen in $Bi_{2}Ru_{2}O_{7}$ is on the similar
basis, although it is considered that such self-doped state would
be higher in metallic $Bi_{2}Ru_{2}O_{7}$ than that in
$Tl_{2}Ru_{2}O_{7}$\cite{Lee3}.

For $Pb_{2}Ru_{2}O_{7-{\delta}}$, both the dc resistivity (see
Fig.1) and the $N_{eff}(\omega_{c})$ deduced from our optical
conductivity spectra show that it is a much better metal than
$Tl_{2}Ru_{2}O_{7}$. Band calculations also show that the band
structures of $Bi_{2}Ru_{2}O_{7}$ and $Pb_{2}Ru_{2}O_{7-{\delta}}$
around $E_{F}$ are quite different from that of
$Tl_{2}Ru_{2}O_{7}$ due to different Ru-O-Ru angles and different
ways of participation of A-cation orbits. The Ru-O-Ru angles in
$Bi_{2}Ru_{2}O_{7}$ and $Pb_{2}Ru_{2}O_{7-{\delta}}$ are larger
than that in $Tl_{2}Ru_{2}O_{7}$, leading to wider bandwidth of Ru
4d band for $Bi_{2}Ru_{2}O_{7}$ and $Pb_{2}Ru_{2}O_{7-{\delta}}$
by changing the corresponding hopping integrals. Since partial Bi
and Pb 6p states situate near $E_{F}$ and mix well with Ru
$t_{2g}$ states around $E_{F}$, the conducting electrons in
$Bi_{2}Ru_{2}O_{7}$ and $Pb_{2}Ru_{2}O_{7-{\delta}}$ are more
itinerant than that in $Tl_{2}Ru_{2}O_{7}$. Therefore, the on-site
Coulomb interaction is relatively weak and the Hubbard bands merge
into a single one\cite{Lee2}. In this case, the knee feature
around ~ 1000 \cm may not be attributed to the interband
transition between LHB and the midgap states for
$Pb_{2}Ru_{2}O_{7-{\delta}}$.

\begin{figure}[t]
\centerline{\includegraphics[width=3.0in]{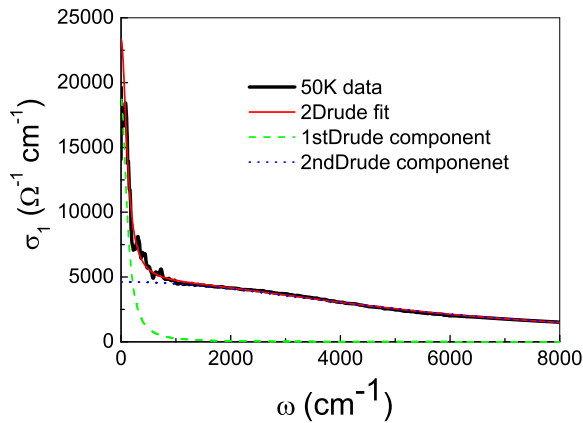}}%
\vspace*{0cm}%
\caption{Fitting with two Drude terms to the optical conductivity
spectrum at 50K.}
\end{figure}

We note that the mid-IR feature for $Pb_{2}Ru_{2}O_{7-{\delta}}$
has its central frequency at zero, suggesting a Drude
characteristic. Then we may consider the low frequency part of the
optical conductivity of $Pb_{2}Ru_{2}O_{7-{\delta}}$ as a sum of
two kinds of intraband transitions, i.e.,
\begin{equation}
 \sigma_{1}(\omega)=\frac{1}{4\pi}\frac{\omega_{p,1}^{2}\cdot\gamma_{1}}{\omega^{2}+\gamma_{1}^{2}}+\frac{1}{4\pi}\frac{\omega_{p,2}^{2}\cdot\gamma_{2}}{\omega^{2}+\gamma_{2}^{2}}
\end{equation}
where $\omega_{p,1}$ and $\omega_{p,2}$ are the plasma
frequencies, $\gamma_{1}$ and $\gamma_{2}$ are relaxation rates of
the free charge carriers. We find that this function can well
reproduce all the spectral data at all temperatures we have measured.
The fitting parameters are listed in table.I.

\begin{figure}[t]
\centerline{\includegraphics[width=3.0in]{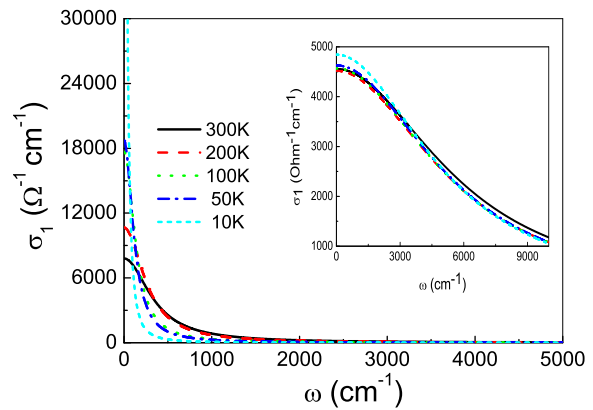}}%
\vspace*{0cm}%
\caption{The temperature dependences of 1st. Drude term and 2nd.
Drude term (in the inset) of the fitting function}
\end{figure}

\begin{table}
\caption{\label{tab:table1}The parameters of function(2)
        at 300 K, 200 K, 100 K, 50 K and 10 K}
\begin{ruledtabular}
\begin{tabular}{ccccc}
 $  T (k)   $&$   \omega_{p,1} (cm^{-1})   $
&$   \gamma_{1} (cm^{-1})   $&$   \omega_{p,2} (cm^{-1})   $&$   \gamma_{2}(cm^{-1})   $\\
\hline
  $  300K   $&$   12779   $&$   349   $&$   40169   $&$   5902   $\\
  $  200K   $&$   13094   $&$   268   $&$   38983   $&$   5605   $\\
  $  100K   $&$   12190   $&$   140   $&$   38975   $&$   5543   $\\
  $  50K    $&$   11616   $&$   120   $&$   39160   $&$   5524   $\\
  $  10K    $&$   12158   $&$   31    $&$   39123   $&$   5268   $\\

\end{tabular}
\end{ruledtabular}
\end{table}

Fig. 4 is the fitting result for spectral data at 50 K, where the
thick solid line is the experimental data, the thin line is a
fitting curve using function(2), the dashed line is the first
Drude term and the dotted line is the second Drude term.

Presented in Fig. 5 is a plot of the fitting curves of the first
Drude term at different temperatures. The inset displays those of
second Drude term. Both of them show usual narrowings with
decreasing temperature. Our fitting results imply that two kinds
of conducting charge carriers with different lifetimes have taken
part in the intraband excitations, and have different temperature
dependencies. The carriers with smaller $\gamma$ value have
stronger temperature dependence, while another one is much weaker.
The cooperation of the two different temperature-dependent
behaviors leads to the decrease of the spectral weight around
$1000cm^{-1}$ in the optical conductivity spectra with decreasing
temperature.

\begin{figure}[t]
\centerline{\includegraphics[width=3.0in]{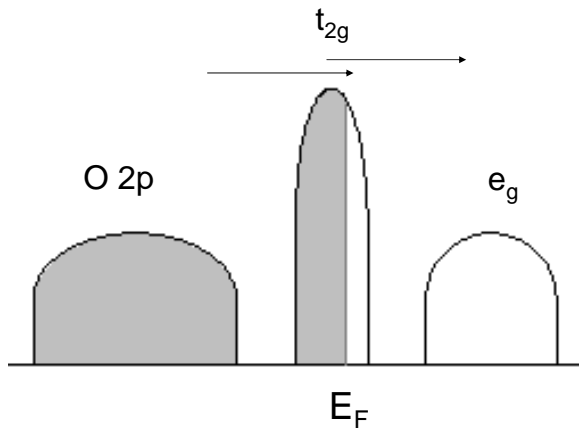}}%
\vspace*{0cm}%
\caption{A schematic picture of electronic states near $E_{F}$ of
$Pb_2Ru_2O_{7-{\delta}}$. The Ru 4d levels are split into $t_{2g}$
and $e_{g}$ bands. The partially occupied $t_{2g}$ bands are
hybridized with partial Pb 6p states. The filled O 2p bands are 2
eV away from the Fermi level.}
\end{figure}

Our analysis is well consistent with the band-structure
calculation results.\cite{Hsu} As displayed in Fig. 6, the
fivefold degenerate Ru 4d levels are split, due to the octahedral
crystal field, into an unoccupied $e_{g}$ band which locate at
~$2\sim5 eV$ above E$_F$, and a partially occupied $t_{2g}$ band
between -1 and 1 eV. This $t_{2g}$ band is broadened by mixing
with some Pb 6p states. The O 2p bands locate between -7.5 eV and
-2 eV.\cite{Hsu} The broad feature at high frequencies (above
$20000 cm^{-1}$) is contributed by two possible interband
transitions with similar energy scales: from the occupied Ru
$t_{2g}$ bands to empty $e_{g}$bands and from the filled O 2p
states to unoccupied Ru $t_{2g}$ states. The low frequency part
below the conductivity minimum is due to the intraband excitations
of the two partially filled Ru $t_{2g}$ bands. The existence of
A-cation orbital near $E_{F}$ actually leads to different band
dispersions of the Ru $t_{2g}$ bands.

This work is in part supported by National Science Foundation of
China and Wang-Kuan-Cheng Foundation for research collaboration
(R.J.). Oak Ridge National Laboratory is managed by UT-Battelle,
LLC, for the U.S. Department of Energy under contract
DE-AC05-00OR22725.

\end{document}